\documentclass[structabstract, rnote]{aa}

\usepackage{amsmath}
\usepackage{url}
\usepackage{txfonts}
\usepackage{natbib}
\usepackage{graphicx}
\bibpunct{(}{)}{;}{a}{}{,} 
\newcommand{\snr}{G296.7--0.9}
\begin{document}

\title{The Supernova Remnant \snr~in X-rays}
\author{T. Prinz\inst{1} \and W. Becker\inst{1,2}}

\institute{Max Planck Institute for extraterrestrial Physics, Giessenbachstrasse, 85741 Garching, Germany \newline\email{tprinz@mpe.mpg.de}
\and Max-Planck Institut f\"ur Radioastronomie, Auf dem H\"ugel 69, 53121 Bonn, Germany}

\date{Received <date> /
Accepted <date>}

\abstract {} {We present a detailed study of the supernova remnant (SNR) \snr~in the 0.2-12 keV X-ray band.}
{Using data from XMM-Newton we performed a spectro-imaging analysis of \snr~in order to deduce the basic 
parameters of the remnant and to search for evidence of a young neutron star associated with it.}
{In X-rays the remnant is characterized by a bright arc located in the south-west direction. Its 
 X-ray spectrum can be best described by an absorbed non-equilibrium collisional plasma model with 
 a column density of $N_\mathrm{H}=1.24_{-0.05}^{+0.07} \times 10^{22}$ cm$^{-2}$ and a plasma 
 temperature of $6.2^{+0.9}_{-0.8}$ million Kelvin. The analysis indicates a remnant age of 5800 to 
 7600 years and a distance of $9.8_{-0.7}^{+1.1}$ kpc. The latter suggests a spatial connection 
 with a close-by HII region. We did not find evidence for the existence of a young neutron star 
 associated with the remnant.}
{}

\keywords{ISM: supernova remnants -  ISM: individual objects: G296.7--0.9 - Stars: neutron}

\maketitle

\section{Introduction}

Every year, several hundred supernova (SN) events are observed, representing the end state
of stellar evolution in which a core-collapse of a massive star or the thermonuclear disruption 
of a white dwarf takes place.  SNe are the most energetic events which can 
be observed in the universe. They are often as bright as a whole galaxy. Their extreme brightness 
allows to see them up to distances of Gpc \citep{2011AAS...21821901R}. At this scale, though, 
the only information obtainable from them is the characteristic rising and fading of their light 
as a function of time, i.e.~their photometric light-curve, and their spectral evolution, both of which 
are important for their classification \citep{2008AJ....135..348S}. Most of these events 
are discovered in the optical band by comparing observations which were taken at different epochs.

In our own Galaxy the rate of observed SNe is small. For core-collapse supernovae it averages 
to only about two per century. This estimate is suggested by the $\gamma$-ray radiation from 
${}^{26}$Al in the Galaxy \citep{2006Natur.439...45D} in which a certain yield is expected to be formed
in each core-collapse SN. Many SNe, though, remain unobserved due to optical light extinction. 
The last recorded SN in our Galaxy, the Kepler SN, was observed in AD 1604 and not more than 
six other SNe have been detected in the Galaxy in the past two thousand years \citep{2003LNP...598....7G}. 
The optical light from the two youngest Galactic SNe, G1.9+0.3 and Cas A, were not observed 
about 100 and 350 years ago \citep{2008ApJ...680L..41R, 2003LNP...598....7G}.

If the direct light from a SN was missed, for several nearby ones we have the chance to 
study at least the light of their remnant, which remains visible in various spectral bands 
for up to $10^5$ years. Although the light echo from the SN Cas A was found recently \citep{2008ApJ...681L..81R}, 
it is one of the few cases so far where the SN light could be studied a few hundred years after the SN. 
The most promising way to learn about the evolution of a SN shock front and the feedback on the evolution 
of their host galaxy is therefore to study the diffuse emission of SNRs. 

In the last years several new SNRs have been detected thanks to the increasing sensitivity of 
modern X-ray observatories. One of those remnants is \snr, which was first detected in X-rays by 
\citet{2002ASPC..271..391S} and later identified to be a SNR by \citet{2003PhDTSchaudel} and 
\citet{2012MNRAS.419.2623R}. It is a shell-like SNR from which X-ray and radio emission has been 
detected. In addition, filaments in the infrared and H$_\alpha$ band were detected in the near 
proximity of the source \citep{2012MNRAS.419.2623R}. Whether they are associated with the 
remnant still remain to be shown.          

A first detailed study of the SNR \snr~was presented by \citet{2012MNRAS.419.2623R} using ROSAT PSPC 
data. As this detector had roughly five independent energy channels in the range 0.1-2 keV, the
spectral results deduced in their analysis were very limited. They suggested that the X-rays were 
emitted from a thermal plasma. \citet{2012MNRAS.419.2623R} therefore were not able to put any 
constraints with high confidence on the derived spectral parameters. The ROSAT data did not allow 
them to deduce parameters like the age or the expansion velocity of the remnant.

In this work we report on an XMM-Newton observation which was targeted on SNR \snr.
The results of the spatial and spectral analysis of this data is presented in 
Section \ref{sec:data}. A discussion is given in Section \ref{sec:discussion} in which we 
use the inferred spectral parameters of \snr~to derive an estimate for its age, its 
radius, its expansion velocity and its distance. Section \ref{sec:summary} provides 
the concluding summary.

\section{X-ray observation and data reduction}\label{sec:data}

\snr~was observed for 13.6 ksec on 28 June 2011 with the EPIC cameras \citep{2001A&A...365L..18S} 
on board the X-ray observatory XMM-Newton (ObsID 0675070101). The two MOS and the PN cameras were 
operated in Full-Frame mode using the medium filter. Part of the observation was taken with the 
filter wheel in closed position because the observation was strongly affected by particle background 
radiation. Therefore, the performed duration was only 4.4 ksec and 5.5 ksec for the PN and MOS1/2 
detectors, respectively.

We used the XMM SAS version 11.0.0 to reprocess and reduce the data. Times of high background
activity were identified by inspecting the light curves of the MOS1/2 and PN data at energies 
above 10 keV. After rejecting these times the effective exposures of the PN, MOS1 and MOS2 cameras 
were 3.7 ksec, 5.1 ksec, and  5.3 ksec, respectively. 
We extracted images and exposure maps in the five standard bands of XMM-Newton using all EPIC instruments. 
Single and double events were selected from the PN data and single to quadruple events from the
MOS data sets. 

For the spectral fitting we used the X-Ray Spectral Fitting Package (XSPEC) version 12.7.0u. 
and restricted the energy range to 
$0.4-6.0$ keV because the count rate detected at higher energies was too sparse for a meaningful 
spectral analysis. Below 0.4 keV the detector and telescope response is not well established. 
All given uncertainties represent the 1$\sigma$ confidence range for one parameter of interest, 
unless stated otherwise.

\subsection{Spatial analysis}\label{sec:spatial}
\begin{figure}
\centering
  \resizebox{0.9\hsize}{!}{\includegraphics[clip]{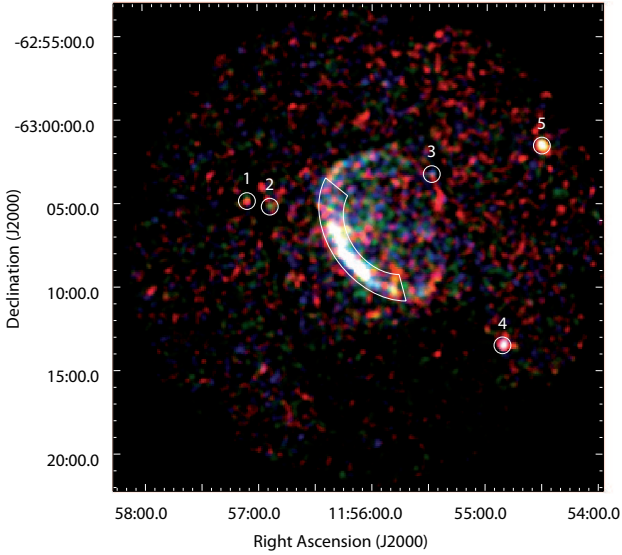}}
  \caption{$30 \arcmin\times 30\arcmin$ XMM-Newton MOS1/2 color image of \snr~(red 
  0.5--0.9 keV, green 0.9--1.3 keV and blue 1.3--2 keV). The superimposed images are 
  binned with $6\arcsec$ per pixel and smoothed by a Gaussian kernel of $\sigma = 30\arcsec$
  to enhance the visibility of the diffuse emission. The photons detected in the annular 
  sector shown in white were used for spectral analysis of the supernova remnant.}\label{fig:G296.7-0.9_rgb}
\end{figure}

Figure \ref{fig:G296.7-0.9_rgb} shows the X-ray color image of SNR \snr. A bright arc 
that appears to lie along an elliptical shell is clearly detected in the 
south-east. Its center is at RA$_\text{c}=11^\text{h}55^\text{m}52\fs3\pm 0\fs7$, 
DEC$_\text{c}=-63\degr06\arcmin29\arcsec\pm 3\arcsec$ with a semi major axis of $5\arcmin$ 
and a semi minor axis of $3\farcm75$. 

We searched for point sources with a signal-to-noise ratio $S/N$ of at least 3$\sigma$ in the five 
standard bands using a sliding box source detection algorithm (SAS tool \sffamily edetect\_chain\normalfont). 
Five point sources were detected (see Fig. \ref{fig:G296.7-0.9_rgb}), which all have a $S/N \ge 7\sigma_G$ 
($\sigma_G=1+\sqrt{c_\text{bg}+0.75}$, where $c_\text{bg}$ are the background counts).
Their position, positional error, signal-to-noise ratio and count rate in the merged image are listed in Table \ref{tab:point_src}. 

\begin{table}
\caption{Detected sources in XMM-Newton observation 0675070101. 
The sources are denoted as in Fig. \ref{fig:G296.7-0.9_rgb}.}
\label{tab:point_src}
\begin{tabular}{ccccccc}
\hline\hline \\[-2ex]
No. 	& RA (J2000) 	& DEC (J2000) 	& S/N & Rate\\ 
	& h:m:s		& d:m:s		& $\sigma_G$ & cts/ksec    \\ \hline\\[-2ex]
1	& $11:57:05.7\pm0.3$	& $-63:04:53\pm2$& 7.7 & 5.2\\
2	& $11:56:53.9\pm0.3$	& $-63:05:13\pm2$& 8.5 & 5.7\\
3	& $11:55:28.6\pm0.3$	& $-63:03:18\pm2$& 7.0 & 4.1\\
4	& $11:54:51.2\pm0.3$	& $-63:13:31\pm2$& 34.1 & 26.4\\
5	& $11:54:31.1\pm0.3$	& $-63:01:33\pm2$& 22.2 & 16.4\\ 
\hline
\end{tabular}
\end{table}

In order to search for optical counterparts we cross-correlated the position of each X-ray source 
with the various online catalogs provided by the VizieR service\footnote{\url{http://vizier.u-strasbg.fr/viz-bin/VizieR}}.
We found a candidate counterpart for source \#5, HD 103442  \citep{2000A&A...355L..27H}, which is an A3/4IV star 
with an optical magnitude $V=8.8$. Its angular separation from source \#5 is $4\farcs0$, corresponding to 
twice the observatory's absolute astrometric accuracy\footnote{\url{xmm.vilspa.esa.es/docs/documents/CAL-TN-0018.pdf}.}. 
The source density $\rho$ in this region of the Galactic plane using the Tycho-2 catalog is $\rho \approx 0.09~\text{arcmin}^{-2}$. 
The probability of a chance association computes then like $P_\text{coin}=\rho\cdot\delta^2$, where $\delta$ 
is the angular separation between the optical and X-ray source. For the optical counterpart of 
source \#5 we compute a chance probability of $4\times 10^{-4}$ for a mis-identification.
Its X-ray flux within the 0.3 to 3.5 keV band is $1.2\times 10^{-12}$ erg cm$^{-2}$
s$^{-1}$, yielding an X-ray to visual flux ratio of $\log(f_\text{X}/f_\text{V})=-3.1$. 
This is in the allowed range of $-3.86 \pm 0.79$ for A-type stars \citep{1988Maccacaro, 2009ApJS..181..444A}
which further strengthens the association between source \#5 and its candidate optical counterpart.

No source has been detected close to the geometrical center of the SNR. 
Using the merged MOS1/2 data we determined a $3\sigma$ upper limit on the count rate 
of a hypothetical central X-ray source of $4.3\times 10^{-3}$ cts/s in the energy 
range 0.2 to 12 keV.

\subsection{Spectral analysis}

The energy spectrum of \snr~was extracted from an elliptical annular sector with 
the center at RA$_\text{c}$ and DEC$_\text{c}$, semi major axes of $5\arcmin$ and
$2\farcm9$, semi minor axes of $3\farcm75$ and $2\farcm2$ and an opening angle 
of $140\degr$. The background spectrum was derived from a nearby region placed on the same MOS1/2 
chip with the same size as the source region. The background contribution was found to be $\approx 50\%$ in 
the two MOS cameras and $\approx 71\%$ in the PN data. After subtracting the background  
1232, 1236, and 3077 source counts remained in the MOS1, MOS2, and PN data. For 
the spectral analysis of the remnant these counts were binned to have at least 75 counts per bin 
in the case of the MOS1/2 observations and 150 counts per bin for the PN data.

We checked whether the spectral fitting results would change if we model the instrumental background 
according to the suggestions made by \citet{2008A&A...478..575K}. They proposed to add a Gaussian 
at the fluorescence line of Al K$\alpha$ at $E=1.49$ keV with zero width and a power law for 
modeling the remaining soft proton contamination, which is not convolved with the 
instrumental response. As it turns out from this analysis the model fits were not significantly 
better than without adding these components using the standard F-test. Thus, the following results 
are without modeling the instrumental background separately.

\begin{figure}
\centering
\resizebox{0.85\hsize}{!}{\includegraphics[angle=-90,clip]{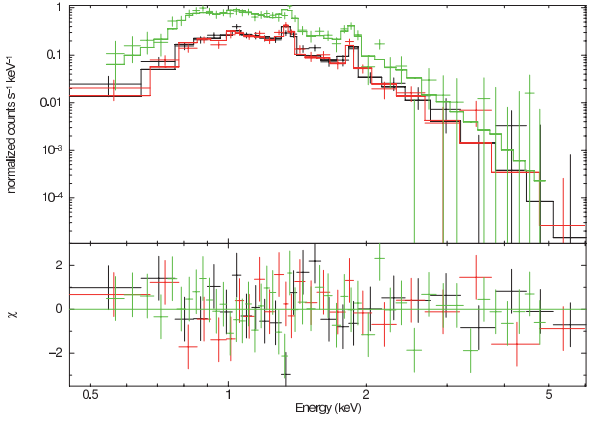}}
\caption{Spectrum and fitted model (\sffamily NEI\normalfont) of the X-ray emission of SNR \snr. 
The black colored line indicates the MOS1 data, the red the MOS2 and the green the PN data.}  
\label{fig:spec_remnant}
\end{figure}

We fitted the spectrum of \snr~with various models: 
A hot diffuse gas model (\sffamily MEKAL\normalfont), a model for a collisionally ionized 
diffuse gas (\sffamily APEC\normalfont), a power law model, a thermal bremsstrahlung model, 
a Raymond-Smith diffuse gas model, a non-equilibrium collisional plasma that allows a 
temperature evolution (\sffamily GNEI\normalfont) or has a constant temperature (\sffamily 
NEI\normalfont, see Figure \ref{fig:spec_remnant}), a plane-parallel shock plasma model (\sffamily PSHOCK\normalfont), 
an ionization equilibrium collisional plasma model (\sffamily EQUIL\normalfont), and a Sedov 
model with separate ion and electron temperatures. For spectral models with $\chi_\text{red}^2<1.4$, 
all fit parameters are listed in Table \ref{tab:spec_param}. 

\begin{table*}
\renewcommand{\arraystretch}{1.2}
\caption{Spectral parameters of the best-fit models for SNR \snr.}
\label{tab:spec_param}
\begin{tabular}{cccccccc}
\hline\hline 
\\[-2.75ex]
model	& $N_\text{H}$ [$10^{22}$cm$^{-2}$]	& $k_\text{B}T$ [keV] & $\tau$\tablefootmark{\alpha} [$10^{11}$ s/cm$^3$]
& $k_\text{B}T_\text{e}$\tablefootmark{\beta}/$\langle kT \rangle$\tablefootmark{\gamma} [keV]& Norm\tablefootmark{\delta} 
[10$^{-2}$ cm$^{-5}$]& $\chi^2$/d.o.f. & $f_X$\tablefootmark{\epsilon} [$10^{-11}$ erg cm$^{-2}$s$^{-1}$]\\ \hline
\sffamily NEI\normalfont& $1.24_{-0.05}^{+0.07}$& $0.53_{-0.07}^{+0.08}$ & $1.2_{-0.4}^{+0.8}$ & -/-			   & $1.3_{-0.3}^{+0.6}$ 
& 89.3/94 & $5.2_{-0.9}^{+1.1}$	\\
\sffamily PSHOCK\normalfont& $1.16_{-0.06}^{+0.14}$& $0.54_{-0.06}^{+0.08}$& $8_{-7}^{+15}$& -/-	& $1.1_{-0.2}^{+0.6}$	
& 87.5/94 & $3.5_{-0.3}^{+0.5}$	\\
\sffamily SEDOV\normalfont& $1.27_{-0.09}^{+0.10}$& $0.35_{-0.08}^{+0.19}$& $6_{-4}^{+7}$& $<0.54$/-	& $2.3_{-0.9}^{+2.1}$	
& 89.4/93 & $8\pm3$	\\
\sffamily GNEI\normalfont& $1.27\pm 0.08$	& $0.49_{-0.08}^{+0.11}$ & $1.0_{-0.5}^{+0.8}$ & -/$0.55_{-0.08}^{+0.11}$ & $1.5_{-0.6}^{+0.9}$ 
& 88.6/93 & $6.0_{-1.5}^{+1.0}$\\
\sffamily EQUIL\normalfont& $1.10 \pm 0.04$	& $0.52\pm 0.04$	& - 		& -/-			& $1.09_{-0.15}^{+0.21}$	
& 96.8/95 & $2.5 \pm 0.2$	\\
\sffamily APEC\normalfont& $1.23\pm 0.05$& $0.38_{-0.03}^{+0.10}$& -	& -/-	& $2.0_{-0.5}^{+0.6}$	
& 107.0/95 & $4.6_{-0.5}^{+0.6}$	\\
\\[-2.75ex] \hline
\end{tabular}
\tablefoot{
\tablefoottext{\alpha}{ionization timescale}
\tablefoottext{\beta}{electron temperature immediately behind the shock front}
\tablefoottext{\gamma}{ionization timescale-averaged temperature 
$\langle kT \rangle=\int_{t_\text{s}}^{t_0}T(t)n_\text{e}(t)\text{dt}/\tau$}
\tablefoottext{\delta}{Norm=$\frac{10^{-14}}{4 \pi [D_\text{A}(1+z)]^2} \int n_\text{e} n_\text{H} 
\text{dV}$, where $D_\text{A}$ is the angular diameter distance to the source in cm, $n_\text{e}$ 
and $n_\text{H}$ are the post-shock electron and hydrogen densities in cm$^{-3}$, respectively.}
\tablefoottext{\epsilon}{X-ray flux in the energy range 0.5 to 4.0 keV.} 
}
\end{table*}

In the following, we will use the best-fit results of the \sffamily NEI \normalfont model to derive related SNR 
parameters. For all models with a 
non-equilibrium equation of state we measure an ionization timescale value 
$\tau=t_0 n_\text{e} <10^{12} \text{ s}\,\text{cm}^{-3}$ ($t_0$ is the age of the remnant and $n_\text{e}$ the post-shock 
electron number density), which is smaller than the expected timescale at which collisional 
ionization equilibrium is reached \citep{2001ApJ...548..820B}. Therefore, the 
\sffamily APEC \normalfont and \sffamily EQUIL \normalfont model are unlikely to apply. 
For the \sffamily GNEI \normalfont model the temperature and the ionization 
timescale-averaged temperature are almost the same, which is no improvement over 
the \sffamily NEI \normalfont model. 
Additionally, we will not use the results of the \sffamily SEDOV \normalfont model any further, 
because the determination of the postshock temperature $T_\text{e}$ in that model is only possible 
at energies above 3 to 4 keV \citep{2001ApJ...548..820B} where the spectrum of \snr~is not well constrained. 

The value for $N_\text{H}$ in the \sffamily NEI \normalfont model fit is slightly lower than the 
average integrated hydrogen column density toward SNR \snr, which is $N_H^{LAB}=1.39\times 10^{22} 
\,\text{cm}^{-2}$ \citep{2005A&A...440..775K}. This value is based on HI emission line measurements at a radio 
frequency of 21 cm and refers to the entire hydrogen column density along the line of sight.
The temperature of the plasma is $6.2^{+0.9}_{-0.8}\times 10^6$ K. No improvement 
in the best-fit statistic was observed when we allowed the abundances to differ 
from the solar values. Using the derived parameters of the \sffamily NEI \normalfont model 
fit the flux in the 0.5 to 4 keV band is $5.2_{-0.9}^{+1.1}\times 10^{-11}$ erg/cm$^2$/s. 
 
In addition we investigated the spatial variation of the spectral parameters by extracting all photons 
in the northern and southern part of the remnant. However, no difference was found within the 
derived errors.

\section{Discussion}\label{sec:discussion}

\subsection{Comparison with the ROSAT results}

\citet{2003PhDTSchaudel} and \citet{2012MNRAS.419.2623R} independently analyzed two pointed ROSAT PSPCB 
observations. These data were taken between 1st and 8th of February 1993 and 19th and 21st 
of February 1998. Both authors found the temperature of the X-ray emitting gas and the hydrogen column 
density $N_\text{H}$ lower by at least a factor of two when compared with the results deduced in our 
work. Only \citet{2003PhDTSchaudel} found a $N_\text{H}$ which is comparable with our value listed 
in Table 2 by fitting the Raymond-Smith model, though his value had a much higher uncertainty. 
As already mentioned in the introduction, the PSPCB had about five independent energy channels, which 
limits the conclusions drawn from their spectral fits, especially when 70 spectral bins were 
used as in the work of \citet{2012MNRAS.419.2623R}. The latter corresponds to an oversampling of about 
15 times the spectral resolution of the detector!

\subsection{Supernova remnant}

Using the deduced spectral parameters from the \sffamily NEI \normalfont fit we can derive
basic properties of the remnant, such as the distance $d$, post-shock hydrogen density $n_\text{H}$,
swept-up mass $M$, the age of the remnant $t$, the radius in pc $R_s$, and the shock velocity $v_s$.

We used the following set of equations, which is described in detail in 
\citet[][]{2012A&A...544A...7P} and references therein.
\begin{eqnarray*}
d_\text{Sedov}&=&7420 \cdot \theta^{-3/5}\left(\frac{E_{51}}{T_\text{s}}\right)^{2/5}
\left(\frac{f}{\text{Norm}}\right)^{1/5}~[\text{kpc}] \\
d_\text{Reddening}&=&N_H/\left( N_\text{H}/A_V\cdot A_V/E_{B-V}\cdot E_{B-V}/\text{kpc}\right)~[\text{kpc}]\\
 V_\text{emit}&=&3.029\times10^{54}\cdot f\cdot \theta^3\cdot (d_\text{Sedov}[\text{kpc}])^3~[\text{cm}^3]\\
n_\text{H}&=&70.27 \sqrt{\frac{\text{Norm}}{d_\text{Sedov}\cdot f\cdot \theta^3}}~[\text{cm}^{-3}]\\
M&=&1.4\cdot  n_\text{H}\cdot  m_\text{H}\cdot  V~[\text{kg}]\\
t&=&2.71\times 10^9 \left(\frac{E_{51}}{n_\text{H}}\right)^\frac{1}{3} 
T_\text{s}^{-\frac{6}{5}}~[\text{yr}]\\
R_\text{s}&=&0.34 \left(\frac{E_{51}}{N_\text{H}}\right)^\frac{1}{5} t^\frac{2}{5}~[\text{pc}]\\
v_\text{s}&=&0.4 \cdot R_\text{s}/t~[\text{km/s}].
\end{eqnarray*}
Herein $\theta=\sqrt{\text{major}\times\text{minor axis}}$ is the reduced radius of the remnant in arcmin, 
$E_{51}$ the explosion energy in units of $10^{51}$ erg, $T_\text{s}$ the fitted plasma temperature, $f$ 
the filling factor to correct for the morphology of the SNR, Norm the normalization of the spectral fit, $N_H$ the fitted  
column density, $V_\text{emit}$ the X-ray emitting volume, and $m_\text{H}$ the mass of a hydrogen atom. 
The errors listed with the numbers deduced from these equations are statistical errors. The systematic
errors might be larger but are unknown.

The Sedov-analysis based distance is $d_\text{Sedov}=9.8_{-0.7}^{+1.1}$ kpc. In the following, we give all 
important quantities in units of $d_{9.8}=d/9.8$ kpc as $d_\text{Sedov}$ has a smaller uncertainty than other
distance estimates.
From Figure \ref{fig:G296.7-0.9_rgb} 
we infer that only $f=15\%$ of the remnant is bright enough to be used for spectral analysis. Therefore, 
the post-shock hydrogen density $N_\text{H}$ is $0.73_{-0.10}^{+0.18}$ and the swept-up mass is $M=29 M_\odot~
d_{9.8}^{5/2}$. Assuming that the explosion energy $E$ is equal to the canonical value of $10^{51}$ erg, the 
age of the remnant is 5800 to 7600 years and the radius is $R_\text{s}=12.2_{-0.8}^{+1.2}~d_{9.8}^{1/6}$ pc. 
We derive a shock velocity $v_s$ of $720_{-100}^{+130}$ km/s. Using the flux values deduced in 
Section \ref{sec:data} we compute its luminosity to be $L_X^{0.5-4}=6.0_{-1.4}^{+1.8}\times 10^{35}~d_{9.8}^{2}$ erg/s. 

Independently of the Sedov analysis another method which allows us to estimate the remnant's distance 
makes use of a relation between the visible extinction $A_\mathrm{V}$ and the color excess: 
$A_\mathrm{V}/E_\mathrm{B-V}=3.2\pm 0.2$. The remnant's color excess 
is not known, so that we use the distribution of the mean color excess $E_\mathrm{B-V}$ per kiloparsec 
derived by \citet{1978A&A....64..367L}. In the direction of the remnant we find $E_{B-V}/\text{kpc}=0.25\pm0.10$. 
In addition, we used the relation between $N_\mathrm{H}$ and the visual extinction $A_\mathrm{V}$ of 
\citet{1995A&A...293..889P} $N_\text{H}/A_V=(1.79\pm 0.03)\times 10^{21}$ cm$^{-2}$. This leads to a 
distance of $d_\text{Reddening}=9\pm 4$ kpc which is in agreement with the distance deduced from the
Sedov analysis. However, the mean color excess $\langle E_\mathrm{B-V}\rangle$ which we used to calculate $d_\text{Reddening}$ 
was derived for a reddening layer up to 2 kpc and thus is just a rough estimate. 

\subsection{Compact central object}

To obtain a rough estimate of the flux upper limit for a hypothetical compact source in the center 
of the remnant, we assumed that the source has properties similar to a Central Compact Object (CCO), 
because no compact source or radio pulsar has been detected in \snr~so far \citep{2012MNRAS.419.2623R}. 
Like other CCOs, e.g. the one in the SNR Puppis A, we assume that 
the spectrum is dominated by blackbody emission with a temperature of $\approx 2.6\times 10^6$ Kelvin 
and a luminosity in the 0.5-10 keV band of at least $10^{32}$ erg/s 
\citep[see][for a review]{2009Becker}. This corresponds to a flux in this energy range of $\approx 9\times 10^{-15}$ 
erg cm$^{-2}$s$^{-1}$. Using the WebPIMMS tool with the fitted $N_\text{H}$ of the SNR the count rate in the 
0.2-12 keV range is $>2\times 10^{-4}$ cts/s for the merged MOS1 and MOS2 observations, an order of magnitude 
lower than the derived $3\sigma$ upper limit for a point-like source at the center of the remnant. 
That we do not detect emission from a CCO in \snr~ does therefore not mean that there is none. The 
observation might not be deep enough to detect it. The type of the supernova is also unconstrained.

\section{Conclusion and summary}\label{sec:summary}

The remnant is characterized by a bright arc in the south-west direction and by diffuse emission with 
low surface brightness in its western part. We showed that the X-ray emission of \snr~is in agreement
with coming from a collisionally heated plasma which has not reached equilibrium yet. The Sedov analysis 
leads to the conclusion that the SNR is about 6600 years old and is expanding with a velocity on the 
order of $\approx 720$ km/s. 

The close-by HII region G296.593--0.975 has a velocity of $+25\pm1 \text{ km s}^{-1}$ based on measured hydrogen 
recombination lines \citep{1987A&A...171..261C}. With the standard IAU parameter for the distance to the center 
of our Galaxy $R_0=8.5$ kpc and the solar orbit velocity of $V_0=220$ km/s derived by \citet{1986MNRAS.221.1023K} 
and the Galactic rotation model of \citet{1989ApJ...342..272F} we deduce a distance to the HII region of 
$9.3\pm0.6$ kpc. For the error estimate we assumed an uncertainty in the velocity-to-distance conversion 
of 7 km/s \citep[e.g.,][]{1988ApJ...333..332C}.

The deduced distance $d_\text{Sedov}$ is in good agreement with the distance of the close-by HII region 
G296.593--0.975. This is a strong indicator for a spatial connection between the SNR and the HII region as 
already suggested by \citet{2012MNRAS.419.2623R}.

The observation used in this analysis was strongly affected by particle background radiation, 
which led to a net observation time that was shorter by a factor of three than the approved exposure 
time. Therefore, only limited statements can be made about the existence of a compact source located 
near the center of the supernova remnant. Deeper observations might help to clarify this question 
and the one of the type of the SN in more detail.

\begin{acknowledgements} 
 We acknowledge the use of the XMM-Newton data archive. T.P. acknowledges support from and 
 participation in the International Max-Planck Research School on Astrophysics at the 
 Ludwig-Maximilians University. 
\end{acknowledgements}

\bibliographystyle{aa} 
\bibliography{literatur} 

\end{document}